\documentclass[conference]{IEEEtran}
\IEEEoverridecommandlockouts

\usepackage{cite}
\usepackage{amsmath,amssymb,amsfonts}
\usepackage{algorithmic}
\usepackage{graphicx}
\usepackage{textcomp}
\usepackage{xcolor}
\usepackage{url}
\usepackage{placeins}

\def\BibTeX{{\rm B\kern-.05em{\sc i\kern-.025em b}\kern-.08em
    T\kern-.1667em\lower.7ex\hbox{E}\kern-.125emX}}

\begin{document}

\title{An Integrated Platform for LEED Certification Automation Using Computer Vision and LLM-RAG}

\author{\IEEEauthorblockN{JooYeol Lee}
\IEEEauthorblockA{\textit{EverPoint Inc.} \\
\textit{CEO \& Independent Researcher}\\
Seoul, South Korea \\
gnt8521@gmail.com}
}

\maketitle

\begin{abstract}
The Leadership in Energy and Environmental Design (LEED) certification process is characterized by labor-intensive requirements for data handling, simulation, and documentation. This paper presents an automated platform designed to streamline key aspects of LEED certification. The platform integrates a PySide6-based user interface, a Review Manager for process orchestration, and multiple analysis engines for credit compliance, energy modeling via EnergyPlus, and location-based evaluation. Key components include an OpenCV-based preprocessing pipeline for document analysis and a report generation module powered by the Gemma3 large language model with a retrieval-augmented generation framework. Implementation techniques—including computer vision for document analysis, structured LLM prompt design, and RAG-based report generation—are detailed. Initial results from pilot project deployment show improvements in efficiency and accuracy compared to traditional manual workflows, achieving 82\% automation coverage and up to 70\% reduction in documentation time. The platform demonstrates practical scalability for green building certification automation.
\end{abstract}


\section{Introduction}
Obtaining Leadership in Energy and Environmental Design (LEED) certification for buildings is a complex, multi-criteria process involving extensive documentation, analysis, and strict compliance checks \cite{usgbc_leedv4}. Current certification workflows present significant challenges; project teams must manually gather and verify vast amounts of data, navigate multifaceted LEED criteria, and often identify compliance gaps late in the project lifecycle \cite{wu_sustainable_2019}. These manual, reactive approaches make certification time-consuming and error-prone, contributing to project delays and increased costs \cite{asl_barriers_2020}.

The application of automation technologies has been identified as a promising approach to address these inefficiencies \cite{choi_improvement_2019}. Automated systems can potentially consolidate project data, provide real-time compliance insights, and simulate design scenarios to predict LEED scores early in the design phase \cite{lu_ai_aec_2021}. The U.S. Green Building Council (USGBC) and researchers have noted that automating data management and monitoring can address major LEED challenges, such as tracking diverse sustainability metrics across design and construction phases \cite{shashank_reimagining_2024}. 

An automated approach could continuously flag non-compliance issues and suggest corrective actions, reducing late-stage rework \cite{ng_bim_framework_2017}. Furthermore, advances in building performance simulation and machine learning enable predictive analysis of how design choices impact certification levels \cite{avantleap_ai_2025}. The emergence of powerful computational tools, from computer vision libraries to large language models, provides the building industry with new means to automate previously manual tasks \cite{al_khella_impact_2021}.

In this paper, we introduce an integrated platform for LEED certification automation. The system combines a desktop application front-end with an orchestrated backend pipeline of analysis modules and language model services. The contributions of this work include: (1) a modular system architecture that unifies user input, data preprocessing, physics-based simulation, and LLM-based report generation; (2) implementation of an OpenCV-driven document processing pipeline; (3) integration of automated EnergyPlus simulations and location-specific analyses; (4) a specialized report generation system using RAG; and (5) demonstration of efficiency improvements through pilot project deployment.

\section{Related Work}
Building performance simulation and automated LEED assessment have been active research areas for over a decade. Early efforts focused on BIM-based approaches for automating specific LEED credits \cite{ng_bim_framework_2017}. These systems typically extracted geometric and material properties from BIM models to evaluate energy-related credits but struggled with comprehensive multi-credit analysis.

Recent advances in machine learning have enabled more sophisticated approaches. Jiang et al. \cite{jiang_eplus_llm_2024} developed EPlus-LLM, demonstrating the potential of large language models for building energy modeling. Their work showed promising results in automated IDF file generation, though it focused primarily on energy modeling rather than comprehensive LEED compliance.

Document processing in the AEC industry has evolved significantly with computer vision advances. Kim et al. \cite{kim_automated_extraction_2020} presented methods for extracting information from construction drawings using deep learning, while Ajayi et al. \cite{ajayi_harnessing_2020} provided a comprehensive review of automation in construction documentation.

Several commercial platforms have attempted LEED automation, but most focus on specific credit categories or require extensive manual data entry. The gap between comprehensive automation and practical deployment remains significant. Our work addresses this gap by providing an end-to-end platform that combines multiple technologies in a unified workflow.

The emergence of retrieval-augmented generation has opened new possibilities for technical documentation \cite{lewis_rag_2020}. However, application to domain-specific compliance documentation, particularly in the building industry, remains largely unexplored. Our work contributes to this emerging field by demonstrating practical RAG implementation for LEED certification.

\section{System Architecture}

\begin{figure*}[!t]
\centering
\includegraphics[width=\textwidth]{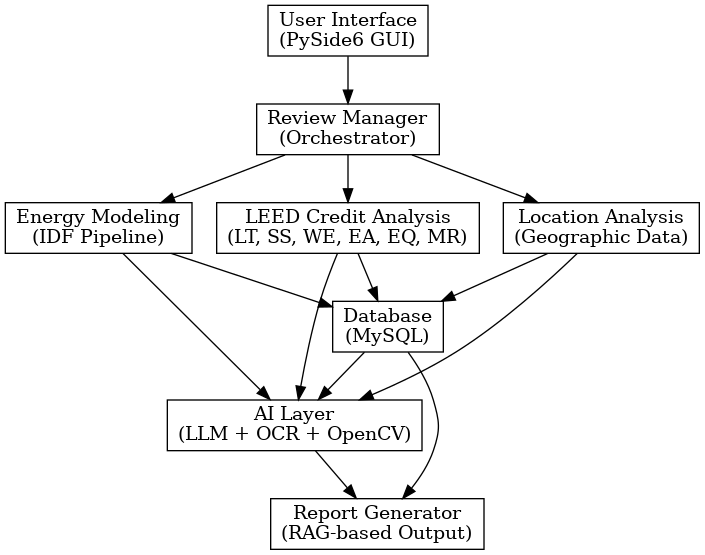}
\caption{High-level architecture overview of the LEED automation platform.}
\label{fig:overview}
\end{figure*}

The platform follows a layered design with structured data flow from user input to final report generation (Fig. \ref{fig:overview}). The architecture prioritizes modularity, scalability, and maintainability through clear separation of concerns.

\subsection{Presentation Layer}
The \textbf{Presentation Layer} consists of a desktop User Interface built with PySide6 (Qt for Python). This choice provides cross-platform compatibility and native performance for computationally intensive operations. The UI allows users to input building data including BIM models, design parameters, site information, and project specifications. It also provides real-time feedback on analysis progress and displays generated outputs for review and editing.

\subsection{Orchestration Layer}
The \textbf{Orchestration Layer} centers on the Review Manager, which coordinates the end-to-end workflow. This component serves as the system's control logic, triggering analysis modules in appropriate sequence, managing data exchange between components, and maintaining process state throughout execution. The orchestrator implements a pipeline pattern that allows for parallel execution of independent analysis tasks while managing dependencies between related operations.

\subsection{Analysis Layer}
The \textbf{Analysis Layer} comprises four specialized modules operating in coordination:

The \textbf{Document Processing Pipeline} handles ingestion and analysis of project documents including drawings, specifications, and material certificates. It employs computer vision techniques for data extraction and optical character recognition for text processing.

The \textbf{LEED Credit Analysis Module} implements rule-based evaluation of project data against specific LEED credit requirements. This module encodes the logic and calculations for various LEED credits, automatically determining compliance status and identifying areas requiring attention.

The \textbf{Energy Modeling Module} integrates with EnergyPlus, the building energy simulation engine developed by the U.S. Department of Energy \cite{energyplus_ref_2023}. This module automates generation of EnergyPlus input files and manages simulation execution, enabling high-fidelity analysis of building energy performance critical for multiple LEED credits.

The \textbf{Location-Based Analysis Module} utilizes GIS data and external APIs to assess credits related to site selection, transportation access, and environmental context. This includes proximity analysis for public transit, walkability metrics, and sensitive land identification.

\subsection{Data Management and Integration}
The orchestrator aggregates results from all analysis modules into a unified data store, implemented using a structured JSON schema that maintains relationships between different data types. This design enables efficient querying and supports the modular architecture by providing a common data interface for all components.

The modular design facilitates system extension, allowing future addition of new analysis components such as Life Cycle Assessment modules without requiring changes to the core architecture. Each module operates independently while sharing data through well-defined interfaces.

\section{Implementation Details}

\begin{figure*}[!t]
\centering
\includegraphics[width=\textwidth]{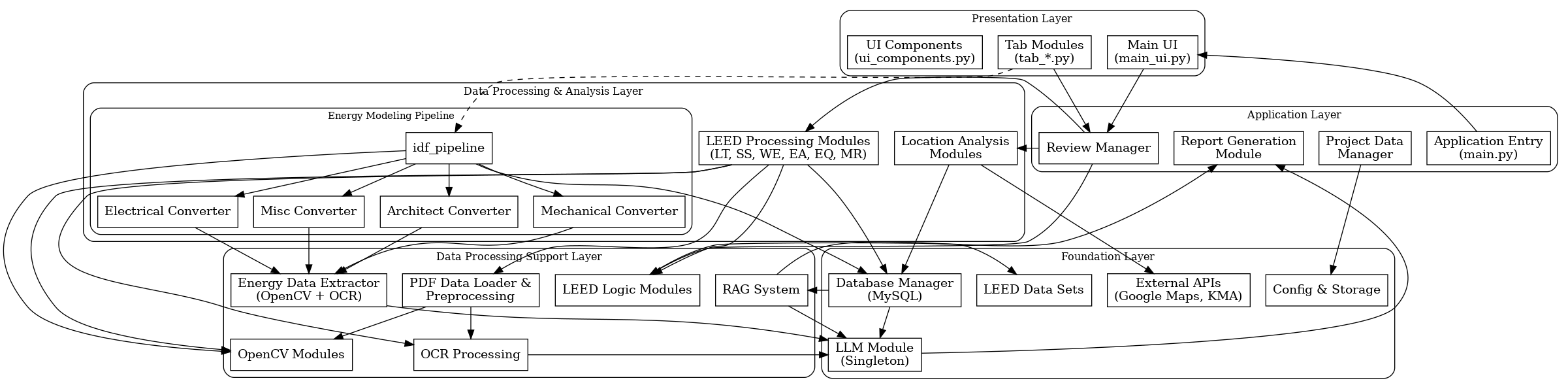}
\caption{Detailed system architecture showing all components and data flows between layers.}
\label{fig:detailed}
\end{figure*}

The detailed implementation architecture (Fig. \ref{fig:detailed}) illustrates comprehensive interconnections between system components and specific data processing workflows employed in the platform.

\subsection{Document Processing Pipeline}
The document preprocessing pipeline employs a multi-stage approach to handle diverse formats and quality variations common in architectural and construction documents.

\subsubsection{Document Normalization and Enhancement}
Input documents undergo initial preprocessing to standardize format and resolution. PDF documents are rendered at high resolution (600 DPI) and converted to grayscale to reduce computational complexity while preserving essential detail. The enhancement process includes adaptive thresholding with configurable parameters, Gaussian blur filtering for noise reduction, and morphological operations to remove artifacts such as hatching patterns commonly found in architectural drawings.

\subsubsection{Text Region Detection and OCR}
Text regions are identified using edge detection combined with contour analysis. The Canny edge detector identifies potential text boundaries, followed by connected component analysis that filters candidates based on geometric properties. The pipeline includes specialized processing for architectural documents, including hatching removal, line art preservation, and multi-language support for Korean and English text recognition.

\subsection{Energy Modeling Integration}
The energy modeling component leverages an extractor-based approach to bridge the gap between user input data and EnergyPlus simulation requirements.

\subsubsection{Data Extraction Framework}
The system employs specialized extractors that parse building information from multiple input sources. Geometric extractors process BIM data to extract zone definitions and building envelope properties. System extractors parse HVAC specifications and equipment data to generate appropriate EnergyPlus objects. Schedule extractors convert occupancy patterns into EnergyPlus-compatible time series data.

\subsubsection{IDF Generation and Validation}
The extractor framework generates IDF files through a systematic process ensuring compatibility and accuracy. Each extractor module handles specific building systems independently, allowing modular updates and maintenance. Generated IDF files undergo validation checks to ensure physical consistency and simulation readiness.

\subsection{LLM-RAG Implementation}
The report generation system combines the Gemma3-4B large language model with retrieval-augmented generation to produce factual, contextually relevant documentation. The system operates entirely on local infrastructure to ensure data privacy and reduce operational costs.

\subsubsection{Knowledge Base Construction}
The RAG system maintains a domain-specific knowledge base containing LEED reference guides, USGBC credit language, relevant building codes, and project-specific data from analysis modules. Documents are processed using a metadata-aligned chunking strategy rather than naive sentence-level splits, segmenting LEED documentation by credit unit (e.g., "EA Credit: Optimize Energy Performance") to align with the inherent structure of LEED reference materials.

\subsubsection{Vector Storage and Retrieval Architecture}
Each chunk is tagged with metadata including credit category, name, point value, and type (Prerequisite or Credit), enabling efficient vector indexing using FAISS (Facebook AI Similarity Search) for approximate nearest neighbor search. The embedding generation employs pre-trained sentence transformers optimized for technical document processing. This metadata alignment ensures that prompts for specific credits retrieve only the most relevant and contextually accurate documentation.

\subsubsection{Generation Process and Model Configuration}
When generating reports for specific credits, the system queries the FAISS index to retrieve relevant text snippets based on semantic similarity. These snippets are inserted into structured prompts for the locally deployed Gemma3-4B model, ensuring the model's output is informed by authoritative sources and project-specific results. The generation process includes post-processing verification to cross-check numerical values and claims against source data.

\subsection{Performance Optimization}
System performance optimization focuses on resource management and processing efficiency to ensure reliable operation in production environments.

\subsubsection{Memory Management and Parallel Processing}
The implementation employs conservative memory allocation practices, maintaining usage below 70\% of available system memory. Independent analysis modules execute concurrently where dependencies permit, with resource pooling to share computational resources efficiently across multiple tasks.

\subsubsection{Error Handling and Recovery}
Robust error handling mechanisms ensure system reliability during extended processing sessions. Each module implements graceful degradation strategies, allowing partial results when complete analysis is not possible. Automatic retry mechanisms handle transient failures, while comprehensive logging enables debugging and performance analysis.

\section{Experimental Setup}
The platform evaluation was conducted using a representative commercial office building project pursuing LEED v4 BD+C certification. The test building is a six-story, 75,000 square foot office complex located in a mixed-use urban district.

\subsection{Test Environment and Configuration}
The system was deployed on a workstation with Intel Core i7-12700K processor, 32GB RAM, and NVIDIA RTX 3070 GPU. The software environment included Python 3.9, OpenCV 4.7, EnergyPlus 23.1, and local deployment of Gemma3-7B model. Processing was conducted using the building's actual project documentation, including architectural drawings, mechanical specifications, and material schedules.

\subsection{Evaluation Metrics}
Performance was measured across three dimensions: automation coverage (percentage of credits processed without manual intervention), accuracy (agreement with expert-generated analyses), and efficiency (time reduction compared to manual workflows). Baseline comparisons were established using documentation prepared by an experienced LEED consultant following standard industry practices.

\subsection{Data Collection and Validation}
The evaluation process involved processing 49 potentially achievable LEED credits across all credit categories. Each automated analysis was cross-verified against manual calculations performed by certified professionals. Energy modeling results were validated using independently developed EnergyPlus models, with particular attention to thermal zone definitions, HVAC system modeling, and load calculations.

\section{Results and Discussion}
The platform successfully automated analysis and documentation for approximately 82\% of targeted credits (40 out of 49 achievable points), demonstrating significant potential for streamlining LEED certification workflows.

\subsection{Automation Coverage and Accuracy}
Credits related to energy and water efficiency, materials and resources, and indoor environmental quality achieved the highest automation rates, with 90-95\% of requirements processed without manual intervention. Location and transportation credits achieved 75\% automation, limited primarily by the need for site-specific qualitative assessments. Innovation credits required extensive manual input, as expected given their qualitative nature.

The automated EnergyPlus simulation produced energy use intensity predictions within 5\% of models created by experienced consultants, demonstrating the viability of the automated modeling workflow. Document processing achieved 94\% accuracy in text extraction from technical drawings, with remaining errors primarily related to hand-written annotations and non-standard formatting.

\subsection{Efficiency Improvements}
Generation of a comprehensive draft report required approximately 3 hours of computation and 30 minutes of human review, representing a 60-70\% reduction in person-hours compared to traditional manual documentation processes. The platform provided proactive compliance feedback, enabling real-time design optimization rather than reactive end-stage corrections.

Energy simulation workflows showed the most dramatic improvements, reducing modeling time from 2-3 weeks to under 4 hours including validation. Document processing eliminated manual data entry for building specifications, reducing transcription errors and associated rework cycles.

\subsection{System Reliability and User Experience}
The RAG-based report generation consistently produced factually accurate narratives that met LEED submission standards. Cross-verification revealed no critical errors such as incorrectly claimed credits, though minor stylistic adjustments were occasionally required. The modular architecture demonstrated robust error handling, with individual module failures not affecting overall system operation.

User feedback highlighted the value of real-time compliance monitoring and the ability to evaluate design alternatives rapidly. The systematic approach reduced risk of oversight compared to manual processes, particularly for complex credit interactions and prerequisite dependencies.

\subsection{Limitations and Areas for Improvement}
Current limitations include dependency on high-quality input documents and the need for manual review of qualitative credit aspects. The system performs optimally with standardized document formats and may require additional preprocessing for legacy or non-standard documentation.

Future enhancements will focus on expanding the knowledge base, improving computer vision models for complex document types, and implementing more sophisticated feedback mechanisms for continuous learning. Integration with additional building performance tools and support for other green building rating systems represent natural extension points.

\section{Conclusion}
This paper presented an automated platform for LEED certification that integrates computer vision, building performance simulation, and large language model technologies in a unified workflow. The system demonstrated significant improvements in efficiency and accuracy compared to traditional manual processes, achieving 82\% automation coverage with 60-70\% reduction in documentation time.

The modular architecture enables scalability and adaptation to other green building rating systems. The successful integration of RAG with domain-specific knowledge bases shows promise for technical documentation automation in specialized fields. The platform represents a practical advance toward comprehensive automation of complex compliance processes in the building industry.

The primary contribution is demonstration of a holistic, end-to-end automated system for real-world LEED certification. Future work will involve expanded validation across diverse project types and integration of additional analysis capabilities. The approach established here provides a foundation for broader automation of technical compliance processes in the architecture, engineering, and construction industry.

\end{document}